\documentclass[conference]{IEEEtran}

\IEEEoverridecommandlockouts                              
\usepackage{graphics}       
\usepackage{epsfig}         
\usepackage{times}          
\usepackage{amsmath}        
\usepackage{amssymb}        
\usepackage{color}
\usepackage{array}

\begin{document}

\title{Lead-Lag-Shaped Interactive Force Estimation by
Equivalent Output Injection of Sliding-Mode}

\author{\IEEEauthorblockN{Michael Ruderman}
\IEEEauthorblockA{University of Agder (UiA), Post box 422,
4604-Kristiansand, Norway \\ Email: \tt\small
michael.ruderman@uia.no}
}

\bstctlcite{references:BSTcontrol}

\maketitle
\thispagestyle{empty}
\pagestyle{empty}

\begin{abstract}
Estimation of interactive forces, which are mostly unavailable for
direct measurement on the interface between a system and its
environment, is an essential task in various motion control
applications. This paper proposes an interactive force estimation
method, based on the well-known equivalent output injection of the
second-order sliding mode. The equivalent output injection is used
to obtain a frequency-unshaped quantity that appears as a matched
external disturbance and encompasses the interactive forces.
Afterwards, a universal lead-lag shaper, depending on dynamics of
the motion control system coupled with its environment, is used to
extract an interactive force quantity. Once identified, the
lead-lag shaper can be applied to the given system structure. An
experimental case study, using a valve-controlled hydraulic
cylinder counteracted by the dynamic load, is demonstrated with an
accurate estimation of the interactive force, that in comparison
to the reference measurement.
\end{abstract}

\section{INTRODUCTION AND BACKGROUND}
\label{sec:1}

Motion control applications are often dealing with weakly known
interactive forces, which directly affect the controlled system
performance and can, in worst case, even provoke instabilities.
The control technologies, where complying forces between the
system and its environment are crucial for a predefined and safe
operation, range from the nanoscale touching devices
\cite{sitti2003} and medical mechatronics
\cite{katsura2005,zemiti2007} to the humanoid-like
\cite{prattichizzo2013,abi2019} and industrial \cite{alcocer2003}
robotics, equally as bulky hydraulic systems
\cite{niksefat2001,koivumaki2015}, here just to refer to some of
them. While structural differences between the motion- and
force-controlled systems and their relationship to mechanical
impedance \cite{hogan1985impedance}, by interaction with
environment, have been highlighted in an elegant way in
\cite{ohnishi1994}, the issues related to coupling of the
interactive forces proved to be challenging. This is especially
when shaping the desired endpoint impedance in the real-world
servo-systems, see e.g. \cite{buerger2007}. Besides, more recent
experimental studies, e.g. \cite{wahrburg2018} in robotics,
demonstrate that an accurate and robust estimation of the contact,
correspondingly external, forces and torques remains a non-trivial
task, even for relatively simple (that case rigid) environmental
couplings and specific tuning of the modeled disturbance dynamics.

\subsection{Interaction with environment}
\label{sec:1:sub:1}

\begin{figure}[!h]
\centering
\includegraphics[width=0.5\columnwidth]{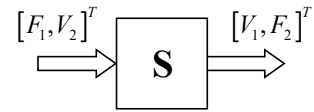}
\caption{Generic two-port of servomechanism with its environment.}
\label{fig:1:1}
\end{figure}

For analyzing couplings of an interactive force, occurring on the
environmental interface, consider a generic two-port
representation $\mathbf{S}$ of the actuated motion system (i.e.
servomechanism) which interacts with its environment, see block
diagram shown in Fig. \ref{fig:1:1}. Recall that the two-port
models, correspondingly networks, with the associated effort
$(F_1,F_2)$ and flow $(V_1, V_2)$ variables, and their product
representing the instantaneous input and output power and, hence,
energy transfer, are particulary useful for modeling interaction
between servomechanisms and environments. That allows specifying a
mechanical impedance and designing an impedance controller
\cite{hogan1985impedance} which, when has a varying closed-loop
stiffness, can be seen as a general form of the motion control
\cite{ohnishi1994}.

Considering, in the most simple case, a linear two-port model of
an interactive motion system (cf. Fig. \ref{fig:1:1}), one can
recognize that the $2 \times 2$ square matrix $\mathbf{S}$ will
contain the transfer functions relating to each other the velocity
and force at each port, cf. \cite{buerger2007}. It is evident that
while $S_{ii}$ describe the transfer characteristics of a
servomechanism and environment, correspondingly, the $S_{ij}$
transfer functions with $i \neq j$ are responsible for the
cross-couplings between both. Assuming $i=1,2$ for the
servomechanism and environment, respectively, and $\mathbf{S}$ to
be regular in terms of invertibility, one can write
\begin{equation}
\left[%
\begin{array}{c}
  F_1 \\
  V_2 \\
\end{array}%
\right] = \mathbf{S}^{-1}
\left[%
\begin{array}{c}
  V_1 \\
  F_2 \\
\end{array}%
\right]
\label{eq:1:1}
\end{equation}
for reverse transfer characteristics of the coupled interactive
system. Introducing $\bar{\mathbf{S}} \equiv \mathbf{S}^{-1}$, one
can recognize that the forces of the servomechanism and
environmental are additionally balanced by the rate of induced
relative motion, meaning $F_1 = \bar{S}_{12} F_2 + \bar{S}_{11}
V_1 $. It is evident that an unconstrained relative motion, i.e.
$\bar{S}_{12} \, \vee \, F_2 = 0$, allows to determine the flow
quantity of servomechanisms from its effort counterpart and vice
versa. As implication, the forward transfer function
$\bar{S}_{11}$ is mostly assumed to be known, correspondingly
identified, for the used nominal servomechanism. On the contrary,
the cross-coupling transfer characteristics $\bar{S}_{12}$ of
environmental interconnection can be barely available and, as
implication, hinder the estimation of external effort variables.
Therefore, an appropriate estimation, or approximation, of the
environmental couplings can be crucial for properly reconstructing
the interactive forces which affect the overall controlled system.

Since an interface between the system and its environment is
application-specific, in the most cases, a suitable reshaping of
the interactive force estimate is required, once the effort
variable $F_2$ is of a primary interest. It is worth noting that
in the most simple case of a directly matched interactive force
(here one can think on an absolutely rigid manipulator hitting a
stiff obstacle with unity restitution coefficient and zero
damping) $\bar{S}_{12}$ will yield the unity or constant transfer
characteristics. On the other hand, when thinking about a standard
solid (also called Zener) model of the viscoelastic type, see e.g.
\cite{tschoegl2012} for fundamentals, one can assume
$$
\bar{S}_{12}(s) = a \, \frac{b \, s + 1}{c \, s + 1},
$$
where $a, \, b, \, c > 0$ coefficients bear the corresponding
elasticity and viscosity constants of the associated environment.
Obviously $s$ is the Laplace variable of the transfer function.
One can recognize that the above transfer function coincides with
the \emph{lead} or \emph{lag} element, for $c < b$ or $c > b$
respectively. With the same line of argumentation, various
structural properties of environmental interfaces, like for
example thermo-rheological, creeping and relaxation, equally as
visco-elasto-plastic effects, can be incorporated into shaping the
external interactive forces. In general, we assume a generic
lead-lag shaper
\begin{equation}
\bar{S}_{12}(s) = a \, \prod_{k=1}^{n} \frac{b_k \, s + 1}{c_k \,
s + 1}, \label{eq:1:2}
\end{equation}
with $a>0$ and $b_k, \, c_k \geq 0$, while the lead-lag order $n
\geq 1$ is the free structural parameter, depending on principles
and mechanisms of the interactive force couplings.

\subsection{Contribution and structure of the paper}
\label{sec:1:sub:2}

This paper is contributing to robust estimation of the interactive
forces, associated with environmental impact and constrained
response, when no explicit parametric modeling of the environment
interface is provided. The principal structural properties of an
interactive force, which is back-propagated to the actuator
dynamics of controlled motion system, are assumed as general
lead-lag characteristics, cf. Section \ref{sec:1:sub:1}. The
corresponding order of the lead-lag shaper is understood to be
rather case-specific, that means depending on the principal
behavior of both, motion control system and its environment. The
proposed method relies on the so-called equivalent output
injection, see e.g.
\cite{edwards2000sliding,davila2006observation} for details, of
the second-order sliding mode
\cite{perruquetti2002sliding,shtessel2014}. Recall that the latter
is robust to the unknown bounded perturbations, has a finite-time
convergence property, and is suitable for using the single output
of second-order systems, for maintaining those in the
sliding-mode. It should also be noted that an equivalent approach,
but involving more detailed explicit modeling of nominal system
dynamics, has been recently shown \cite{ruderman2019} for the same
experimental data.

Following assumptions are made for the rest of the paper. (i) a
time-continuous system dynamics is uniformly considered, despite
all real-time implementations are using the forward Euler
discretization scheme\footnotemark{\footnotetext{Assumption (i) is
justified by the sampling time of 1 millisecond -- twice smaller
in the order of magnitude than the time constants of the system
demonstrated in the experimental case study of this work.}}. (ii)
initial conditions are negligible so that the transient phases,
equally as convergence phase to the sliding-mode, are taken out
evaluation, correspondingly performance assessment. (iii) neither
noise by-effects nor sliding-mode related chattering are within
the scope of the recent work and, therefore, neglected in both the
analysis and experimental evaluation. (iv) for the sake of
generality, especially in relation to a robust shaper design in
frequency-domain and lack of an accurate friction identification
(see e.g. \cite{ruderman2015observer} for more details on
frictional uncertainties) the dynamic friction effects are taken
out of consideration.

The main content of the paper is organized as follows. In Section
\ref{sec:2} the second-order sliding mode, correspondingly the
associated exact differentiator, are summarized for the sake of
clarity. An optimal parameter setting, according to
\cite{VenturaFridman2019}, is briefly addressed. The proposed
estimation of interactive forces is described in Section
\ref{sec:3}, together with the corresponding lead-lag shaping of
equivalent output injection. An experimental case study, dedicated
to predicting the interactive load forces in a controlled
hydraulic cylinder system, is provided in Section \ref{sec:4}. The
paper is concluded by Section \ref{sec:5}.

\section{SECOND-ORDER SLIDING MODE}
\label{sec:2}

The so-called second-order sliding mode, see e.g.
\cite{perruquetti2002sliding,shtessel2014} for fundamentals,
appears when a sliding variable $\sigma$ satisfies
\begin{equation}
\sigma = \dot{\sigma} = 0,
\label{eq:2:1}
\end{equation}
while $\sigma = \sigma(t,\mathbf{x}) \in \mathbb{R}$ is a
sufficiently smooth function of time $t$ and system states
$\bold{x}$, and understood in the Filippov sense
\cite{filippov1988}. The main issue with using higher (than first)
order sliding modes is the demand on system states to be
available, correspondingly
measurable\footnotemark{\footnotetext{This is excluding the
approaches where the high-order sliding-mode (HOSM)
differentiators \cite{levant2003higher,Moreno2018} are used for
reconstructing the dynamic system states from the given single
output measurement.}}. This means for fulfilling \eqref{eq:2:1},
both $\sigma$ and $\dot{\sigma}$ should be determinable as from
the system states, cf. with Chapter 3 in
\cite{perruquetti2002sliding}. Single exception is the well-known
super-twisting algorithm (STA) \cite{levant1993sliding} which
needs the measurement of $\sigma$ only, for steering the system
into the second-order sliding mode. STA drives both $\sigma, \,
\dot{\sigma} \rightarrow 0$ in finite time, so that a second-order
sliding mode occurs after the system reaches the globally stable
origin $(\sigma, \, \dot{\sigma})=\mathbf{0}$.

Based thereupon, the first-order robust differentiator, introduced
by Levant in \cite{levant1998robust}, can be written as
\begin{eqnarray}
\label{eq:2:2}
\dot{\hat{x}}_1 &=& K_1 \sqrt{|e|} \, |e|^{-1} e + \hat{x}_2, \\
\dot{\hat{x}}_2 &=& K_2 \, |e|^{-1} e. \label{eq:2:3}
\end{eqnarray}
It aims at providing an exact estimation of unavailable $
\dot{\sigma}(t>T) \equiv \hat{x}_2$ quantity, after a finite
convergence time $T>0$. The estimator dynamics, given by
\eqref{eq:2:2}, \eqref{eq:2:3}, is driven by the output error
$e=\sigma -\hat{x}_1$, while only the sliding variable $\sigma(t)$
is available from the system measurements. For the appropriately
chosen estimator gains $K_1, K_2 > 0$, which are the STA
parameters \cite{levant1998robust}, the robust exact
differentiator ensures convergence of the states estimation, i.e.
$e = \dot{e} = 0$, and that after finite-time transients. This is
generally valid for an upper bounded second-order dynamics, where
$|\ddot{\sigma}| \leq L = \mathrm{const} < \infty$ denotes the
Lipschitz constant to be known. The positive constant $L$ is
understood to upper bound the matched, but unknown, disturbances
of the nominal second-order dynamics.

For an optimal STA gain setting, one can assume
\begin{equation}
K_1 = 2.028 \sqrt{K_2}, \quad  K_2 = 1.1 L, \label{eq:2:4}
\end{equation}
as has been described and analyzed in detail in
\cite{VenturaFridman2019}. Here is is worth noting that the STA
gain setting \eqref{eq:2:4} aims for minimizing the amplitude of
fast oscillations, i.e. amplitude of chattering, in the
closed-loop of STA estimator. Further, one can notice that the
above $K_2$-selection, with respect to $L$, is the standard one,
also for the HOSM derivatives, as initially proposed in
\cite{levant1998robust} and later confirmed in multiple works, see
e.g.
\cite{levant2003higher,reichhartinger2018arbitrary,VenturaFridman2019}.
The optimal gain setting \eqref{eq:2:4} has also been recently
evaluated with experiments in \cite{RudFrid2018}. From the above,
it is obvious that an appropriate gains assignment requires the
upper bound of the disturbed second-order dynamics to be known.
This is a well-known and studied issue when designing the
STA-based estimators, equally as control algorithms, see e.g.
\cite{levant2007principles}. If $L$ is unavailable from some
nominal system description, correspondingly design, its
approximative estimation is to be obtained based on the
experimental data. An example of such identification approach
aimed for determining $L$ is shown further on in Section
\ref{sec:4}, within the provided experimental case study.

\section{ESTIMATION OF INTERACTIVE FORCES}
\label{sec:3}

For estimating the interactive forces of environment, consider a
perturbed second-order dynamic system as
\begin{equation}
\ddot{\sigma} = f(u, \sigma, \dot{\sigma}, t) + \xi(t).
\label{eq:3:1}
\end{equation}
The unperturbed (nominal) system dynamics is captured by
$f(\cdot)$, including the linear scaling factor of the inertial
mass $m$. The most simple case, of an actuated unconstrained
motion\footnotemark{\footnotetext{The case is considered in the
experimental study provided in Section \ref{sec:4}.}}, one assumes
$f = m^{-1}\bigl(u-d(\dot{\sigma})\bigr)$ where an available input
value is equivalent to the controlled force of the servomechanism,
i.e. $u \equiv F_1$. The induced motion dynamics is counteracted
by the velocity-dependent damping $d(\cdot)$, that is (mostly) the
Coulomb and/or viscous friction, both inherent for the moving
bodies with bearings, correspondingly contact surfaces, of an
actuated relative displacement. For a controlled servomechanism
coupled with its environment, the interactive forces are provoking
an unknown, yet upper bounded, perturbation $\xi(t)$. The
boundedness assumption of the perturbation dynamics follows
directly from the naturally limited interactive forces, for which
$|F_2| < F_{\max}$ is guaranteed for the finite system
accelerations, input excitations, and some constant $F_{\max}$.
The boundedness assumption argues again in favor of the lead-lag
shaped couplings with environment, cf. \eqref{eq:1:2}, meaning it
excludes the free integrators or differentiators when determining
$\bar{S}_{12}$. We also stress that due to the boundedness
assumption of the perturbed second-order dynamics, the
second-order sliding mode appears particulary suitable for a
robust estimation of the unknown interactive forces.

For the perturbed case of an exact differentiator \eqref{eq:2:2},
\eqref{eq:2:3} we introduce the state estimation error
$\tilde{x}_2 = \dot{\sigma} - \hat{x}_2$ which dynamics is,
consequently, governed by
\begin{equation}
\dot{\tilde{x}}_2 = f(\cdot) + \xi(t) - K_2 \, |e|^{-1} e.
\label{eq:3:2}
\end{equation}
Note that the nominal system dynamics, here and in the following,
is written without explicit arguments, this for the sake of
simplicity and for not forcing oneself to have time-varying and
full-state-dependent dynamics. The finite-time convergence to the
second-order sliding mode set ensures that there exists a time
constant $T > 0$ such that for all $t \geq T$ the following
identity holds $0 \equiv \dot{\tilde{x}}_2$
\cite{davila2006observation}, thus leading to
\begin{equation}
K_2 \, |e|^{-1} e = f(\cdot) + \xi(t). \label{eq:3:3}
\end{equation}
Thereupon, an equivalent output injection, cf.
\cite{davila2006observation}, is
\begin{equation}
\chi \equiv K_2 \, \mathrm{sign}(e) = f(\cdot) + \xi(t).
\label{eq:3:3}
\end{equation}
Theoretically, an equivalent output injection is determined by an
infinite switching frequency of the discontinuous term, which is
maintaining the converged second-order sliding mode. It implies
that the spectral distribution of equivalent output injection
contains both, the known part of the motion dynamics and unknown
coupled interactive forces, in addition to high-frequent
oscillations of the sliding-mode known as chattering
\cite{perruquetti2002sliding,shtessel2014}. Since the practical
finite-sampling of an estimator (in original work
\cite{davila2006observation} also called \emph{observer}) produces
a high but finite switching frequency, the necessity to apply a
filter to $\chi$ becomes self-evident. Most simple case, a finite
impulse response (FIR) unity gain low-pass filter, denoted by $h$,
can be designed in frequency domain and used as a chattering
cut-off operator. This, rather standard \cite{shtessel2014},
filtering approach that allows using an equivalent output
injection, will indispensably provoke an additional phase lag in
the estimate
$$
\hat{\xi}(t) \equiv h \bigl[\chi(t) - f(\cdot)(t)\bigr] = \xi(t) +
\varepsilon(t).
$$
Here $\varepsilon(t)$ is the dynamic perturbation difference
caused by the filtering process, while $\varepsilon(s) \rightarrow
0 $ for $\omega \rightarrow 0$, for $\omega$ to be the angular
frequency. Therefore, the filtering by $h$ causes no errors in the
lower frequencies.

Instead of low-pass filtering the equivalent output injection, we
make use of the lead-lag transfer characteristics of the
environmental couplings, cf. Section \ref{sec:1:sub:1}. Without
loss of generality and needs of specifying the polynomial
coefficients and order of \eqref{eq:1:2}, we can distinguish two
principally different classes of environmental interfaces -- of
the lead- or of the lag-type at higher angular frequencies
$\omega$. While both will approach the $a$-gain at steady-state,
i.e. for $\omega \rightarrow 0$, an application-specific finite
gain enhancement will be otherwise expected for the lead-type
interfaces at $\omega \rightarrow \infty$. Consequently logical, a
lag-type environmental interface will exhibit a finite
gain-reduction at high frequencies, i.e. at $\omega \rightarrow
\infty$. Falling back on a viscoelasticity type interface
modeling, as explained in Section \ref{sec:1:sub:1}, some general
remarks can be drawn to attention. If, during the principal
behavior of environmental interface, the elasticity will be
dominating over viscosity, a lag-type coupling of the interactive
forces can be expected. On the contrary, a lead-type environmental
coupling is to be expected when the viscosity effects on the
interface dominate over elasticities in the structure. One should
keep in mind that the above distinguishing between the lead- and
lag-type interfaces refer to an upper bound of the excitation
frequencies. At the same time, an application-specific shaping of
the overall transfer characteristics of the coupling interface is
required for $0 < \omega < \infty$, thus giving reasoning to the
generic shaper \eqref{eq:1:2}.

The above considerations allow for using the lead-lag shaper and,
with the introduced transfer function $G(s) \equiv m \,
\bar{S}_{12}^{-1} (s)$, designing an infinite impulse response
(IIR) filter $g(\cdot)$, which is the inverse Laplace transform of
$G$. It is worth emphasizing that the transfer characteristics,
captured by $G$, do not reflect an (artificially) injected
low-pass filter, but have a direct relationship to the coupling
interface properties of the system $\mathbf{S}$, cf. Section
\ref{sec:1:sub:1}. Hence, the estimated interactive force can be
obtained from the equivalent output injection as
\begin{equation}
\hat{F}_2(t) = g \bigl(K_2 \, \mathrm{sign}(e)(t) - f(\cdot)(t)
\bigr). \label{eq:3:4}
\end{equation}

An essential point, to be equally mentioned here, is that an
unavailable system state can enter the nominal dynamics
$f(\cdot)$. This case, the state estimate, e.g. $\hat{x}_2$, has
to be used instead of the unmeasurable system quantity. Yet this
leads to a feedback-coupled estimator dynamics and, as a logical
consequence, to an additional initial perturbations
$f(\hat{x}_2)(t) - f(\dot{\sigma})(t)$ for $t < T$, i.e. before
the finite-time convergence of the robust differentiator, cf.
Section \ref{sec:2}. Still, when fairly requiring the boundedness
of an initial state discrepancy and BIBO characteristics of the
nominal dynamic map $f(\cdot)$, one can neglect the transient
phase $t < T$ and assumes $f(\hat{x}_2) \approx f(\dot{\sigma})$
$\forall \, t>T$, i.e. once the system is in sliding-mode. For the
related convergence analysis and observer stability, in spite of a
feedback-coupled estimation dynamics, an interested reader is
referred to \cite{davila2005second}.

\section{EXPERIMENTAL CASE STUDY}
\label{sec:4}

The experimental case study is accomplished on a valve-controlled
hydraulic cylinder system, counteracted by another cylinder which
appears as a dynamic system load. Both cylinders are rigidly
coupled to each other via a sensing force-cell, that allows for
direct reference measurement of the interactive forces which we
aim to estimate, correspondingly predict. More technical details
on the experimental setup of hydraulic system in use can be found
in \cite{PasolRuder2018,PasolRuder2019}.

The single system parameter identified prior to the experimental
study is the overall lumped moving mass $m$, which appears as a
scaling factor in the total force balance. Both, the shaping
lead-lag dynamics
\begin{equation}
G(s) = 1.7 \, \frac{2.84 \times 10^{-5} \,s + 1}{0.00137 \, s + 1}
\cdot \frac{2.38 \times 10^{-5} \,s + 1}{0.01284 \, s + 1},
\label{eq:4:1}
\end{equation}
and the unknown Coulomb friction coefficient $\gamma$, resulting
in
\begin{equation}
f=m^{-1} \bigl(u - \gamma \, \mathrm{sign}(\hat{x}_2) \bigr) =
0.5882 \bigl(u - 160 \, \mathrm{sign} (\hat{x}_2) \bigr),
\label{eq:4:2}
\end{equation}
are identified simultaneously, by a standard numerical
minimization routine, using the measured reference force data.

\begin{figure}[!h]
\centering
\includegraphics[width=0.95\columnwidth]{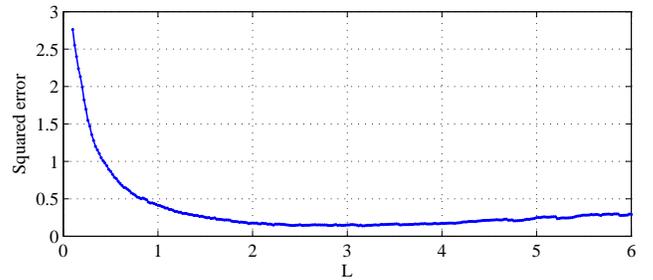}
\caption{Cumulative squared error against varying $L$.}
\label{fig:4:0}
\end{figure}

Since $L$ remains the single unknown design parameter of the
STA-based estimator, the proposed approach aims at determining it
via numerical optimization. That is performed on experimental data
of the single measured output. Here it is worth to recall that the
reference force measurement can be unavailable during the design
stage. Solving minimization
\begin{equation}
\underset {L} \min \sum \limits_{i=1}^{N} e(L)^2, \label{eq:4:3}
\end{equation}
of the squared output error yields $L=3.1$. Here $N$ is the size
of the measured and STA-estimated data, while the output error $e$
depends on the $L$-assignment affecting the STA-gains cf.
\eqref{eq:2:4}. The cumulative squared error \eqref{eq:4:3} is
shown in Fig. \ref{fig:4:0} against the varying $L$, out of which
an optimal $L$-value is read off.

\begin{figure}[!h]
\centering
\includegraphics[width=0.95\columnwidth]{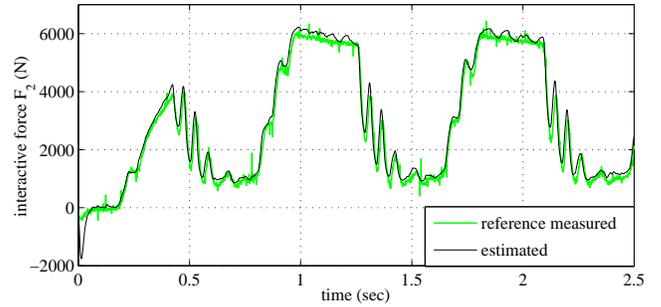}
\caption{Estimated interactive force $\hat{F}_2$ versus reference
measured $F_2$.} \label{fig:41}
\end{figure}

\begin{figure}[!h]
\centering
\includegraphics[width=0.95\columnwidth]{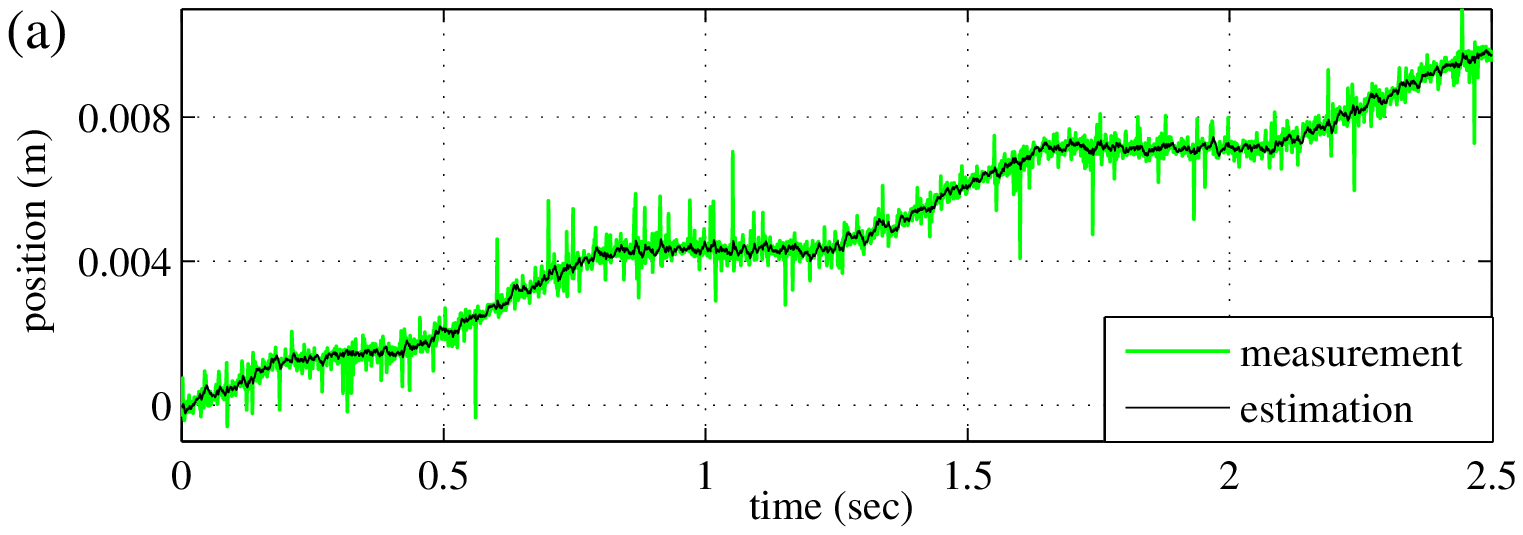}
\includegraphics[width=0.95\columnwidth]{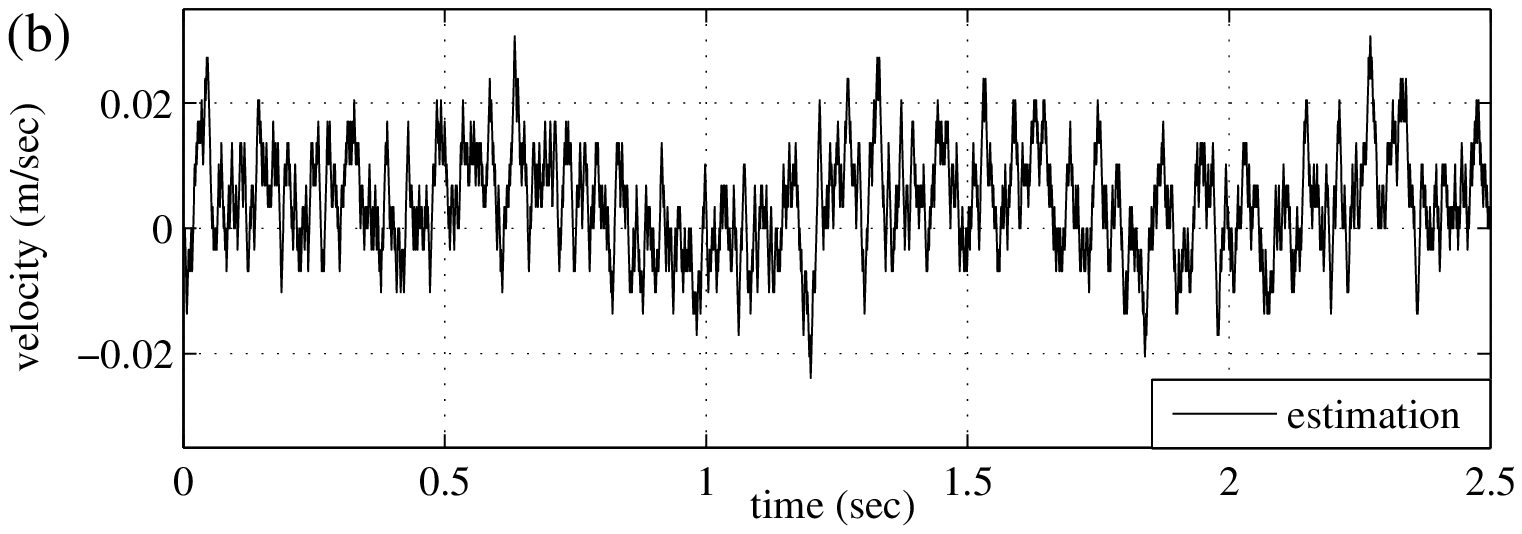}
\caption{Trajectories of induced motion with interactive force,
measured versus estimated relative displacement (a) and estimated
relative velocity (b)} \label{fig:42}
\end{figure}

The reference measured interactive force, used for the above
parameters identification, is shown versus the estimated one in
Fig. \ref{fig:41}. One can recognize both time series are well in
accord with each other, and that for transient, oscillating, and
quasi steady-state values of lower (about 1000 N) and higher
(about 6000 N) amplitudes.
The corresponding motion profile, with the measured and estimated
quantities of relative displacement and velocity, are shown in
Fig. \ref{fig:42} (a) and (b) respectively. One can recognize a
relatively high level of the displacement measurement noise which
indirectly argues in favor of the robust sliding-mode-based
estimation scheme. From Fig. \ref{fig:42} (b) one can further
recognize, that the relative motion is with relatively low
velocity amplitudes. The velocity pattern is frequently
oscillating in a stick-slip manner, also with multiple sporadic
zero-crossings, that is typical for slower displacements under
impact of a high process noise and external perturbations, cf.
Fig. \ref{fig:42} (a).

\begin{figure}[!h]
\centering
\includegraphics[width=0.95\columnwidth]{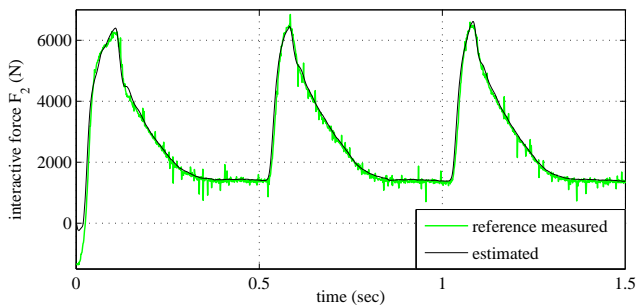}
\caption{Estimated interactive force $\hat{F}_2$ versus reference
measured $F_2$.} \label{fig:43}
\end{figure}

Another set of unseen data, i.e. not involved into parameters
identification, has been equally used for evaluating the
estimation of an interactive force. Here the estimated and
reference measured interactive force values are shown opposite to
each other in Fig. \ref{fig:43}.
This time, the interactive force has more steeply periodic peaks,
coming from the saw-shaped profile of the applied counteracting
load, and the longer steady-state plateaus in-between, cf. Fig.
\ref{fig:43}. Also here one can recognize a good accord between
the estimation and measurement.

\section{CONCLUSIONS}
\label{sec:5}

For robust estimation of the unknown interactive forces, a method
based of the second-order sliding-mode and associated equivalent
output injection principles has been proposed. It is shown that,
depending on the system dynamics and interactive force couplings,
which appear as matched perturbations, the equivalent output
injection can be reshaped via the standard lead-lag transfer
characteristics. Design of the estimation method is presented
along with the parametrization of an exact differentiator
\cite{levant1998robust} and the proposed strategy of reshaping the
equivalent output injection quantity. An experimental case study,
showing an accurate estimation of the interactive force in the
dynamically loaded valve-controlled hydraulic cylinder with high
measurement and process noise, is provided for evaluation of the
proposed method.

\section*{Acknowledgment}
This work has received funding from the European Union Horizon
2020 research and innovation programme H2020-MSCA-RISE-2016 under
the grant agreement No 734832. Discussions with Prof. Leonid
Fridman on equivalent output injection principles are further
gratefully acknowledged.

\bibliographystyle{IEEEtran}
\bibliography{references}

\end{document}